\documentclass[twocolumn,showpacs,preprintnumbers,superscriptaddress,amsmath,amssymb,floatfix]{revtex4}
\usepackage{graphicx}% Include figure files
\usepackage{dcolumn}% Align table columns on decimal point
\usepackage{bm}% bold math
\usepackage{color}
\usepackage{xcolor}
\usepackage{amssymb}
\usepackage{amsmath}
\usepackage{mhchem}
\usepackage{revsymb}
\usepackage{gensymb}
\begin{document}

\preprint{APS/123-QED}

\title{Lifetime measurements of excited states in $^{15}$O}

\author{B.~Frentz}
% \email[Electronic Address: ]{bfrentz@nd.edu}
\author{A.~Aprahamian}
\author{A.M.~Clark}
\author{C.~Dulal}
\author{J.D.~Enright}
\author{R.J.~deBoer}
\author{J.~G\"{o}rres}
\author{S.~L.~Henderson}
\author{K.B.~Howard}
\author{R.~Kelmar}
\author{K.~Lee}
\author{L.~Morales}
\author{S.~Moylan}
\author{Z.~Raman}
\author{W.~Tan}
\author{L.~E.~Weghorn}
\author{M.~Wiescher}
\affiliation{Department of Physics, University of Notre Dame, Notre Dame, Indiana 46556 USA}
\affiliation{The Joint Institute for Nuclear Astrophysics, University of Notre Dame, Notre Dame, Indiana 46556, USA}
%\affiliation{Department of Physics, University of Notre Dame, Notre Dame, Indiana 46556 USA}
%\affiliation{The Joint Institute for Nuclear Astrophysics, University of Notre Dame, Notre Dame, Indiana 46556, USA}

\date{\today}% It is always \today, today,
             %  but any date may be explicitly specified

\begin{abstract}

The CNO cycle is the main energy source in stars more massive than our sun, it defines the energy production and the cycle time that lead to the lifetime of massive stars, and it is an important tool for the determination of the age of globular clusters. One of the largest uncertainties in the CNO chain of reactions comes from the uncertainty in the $^{14}$N$(p,\gamma)^{15}$O reaction rate. This uncertainty arises predominantly from the uncertainty in the lifetime of the sub-threshold state in $^{15}$O at $E_{x}$ = 6792 keV. Previous measurements of this state's lifetime are significantly discrepant. Here, we report on a new lifetime measurement of this state, as well as the excited states in $^{15}$O at $E_{x}$ = 5181 keV and $E_{x}$ = 6172 keV, via the $^{14}$N$(p,\gamma)^{15}$O reaction at proton energies of $E_{p} = 1020$ keV and $E_{p} = 1570$ keV. The lifetimes have been determined with the Doppler-Shift Attenuation Method (DSAM) with three separate, nitrogen-implanted targets with Mo, Ta, and W backing. We obtained lifetimes from the weighted average of the three measurements, allowing us to account for systematic differences between the backing materials. For the 6792 keV state, we obtained a $\tau = 0.6 \pm 0.4$ fs. To provide cross-validation of our method, we measured the known lifetimes of the states at 5181 keV and 6172 keV to be $\tau = 7.5 \pm 3.0$ and $\tau = 0.7 \pm 0.5$ fs, respectively, which are in good agreement with previous measurements.

\end{abstract}

\pacs{Valid PACS appear here}% PACS, the Physics and Astronomy
                             % Classification Scheme.
%\keywords{Suggested keywords}%Use showkeys class option if keyword
                              %display desired
\maketitle

\section{Introduction}

The measurements of solar neutrinos \cite{McDonald_2004} have confirmed the prediction of the $pp$-chains as the dominant energy source of our sun. Only about 1\% of the solar energy in the sun comes from the CN cycle, 
\vspace{-10pt}

\begin{align*}
\ce{^{12}C} \left( p, \gamma \right) \ce{^{13}N} \\
\ce{^{13}N} \left( \beta^{+} \nu \right) \ce{^{13}C} \\
\ce{^{13}C} \left( p, \gamma \right) \ce{^{14}N} \\
\ce{^{14}N} \left( p, \gamma \right) \ce{^{15}O} \\
\ce{^{15}O} \left( \beta^{+} \nu \right) \ce{^{15}N} \\
\ce{^{15}N} \left( p, \alpha \right) \ce{^{12}C},
\end{align*}

\noindent but it quickly becomes the dominate source of energy production in stars with masses $M \gtrsim M_{\odot}$. One of the major uncertainties in the description of the sun, in the framework of the standard solar model \cite{PhysRevLett.92.121301,Bahcall_2006}, is the metallicity of the solar core, which is determined by its carbon, nitrogen, and oxygen content \cite{ASPLUND20061}. The expected element abundances, based on the spectroscopic analysis of the solar atmosphere, disagree with the solar profiles of sound speed and density as well as the depth of the convective zone and the helium abundance obtained by helioseismology measurements \cite{Bahcall_2005}. It has been pointed out that a direct study of the CN neutrinos, coming from the $\beta$ decay of $^{13}$N and $^{15}$O, can provide an independent measure of the solar metallicity \cite{Haxton_2008}. However, the CN neutrino flux not only depends on the CN abundance in the solar interior, but also on the associated CN reaction rates, such as $^{12}$C($p,\gamma$)$^{13}$N and $^{14}$N($p,\gamma$)$^{15}$O respectively.

The BOREXINO collaboration has succeeded in the first measurement of the CNO neutrinos associated primarily with the $^{15}$O $\beta$ decay \cite{agostini2020direct}. The abundance of $^{15}$O depends critically on the $^{14}$N$(p,\gamma)^{15}$O reaction rate, which is the slowest in the CN cycle, and therefore determines the $^{15}$O equilibrium abundance. Extensive experimental efforts have been undertaken to determine a reliable reaction rate at the solar energy range.  

\begin{figure}
\includegraphics[width=1.0\columnwidth]{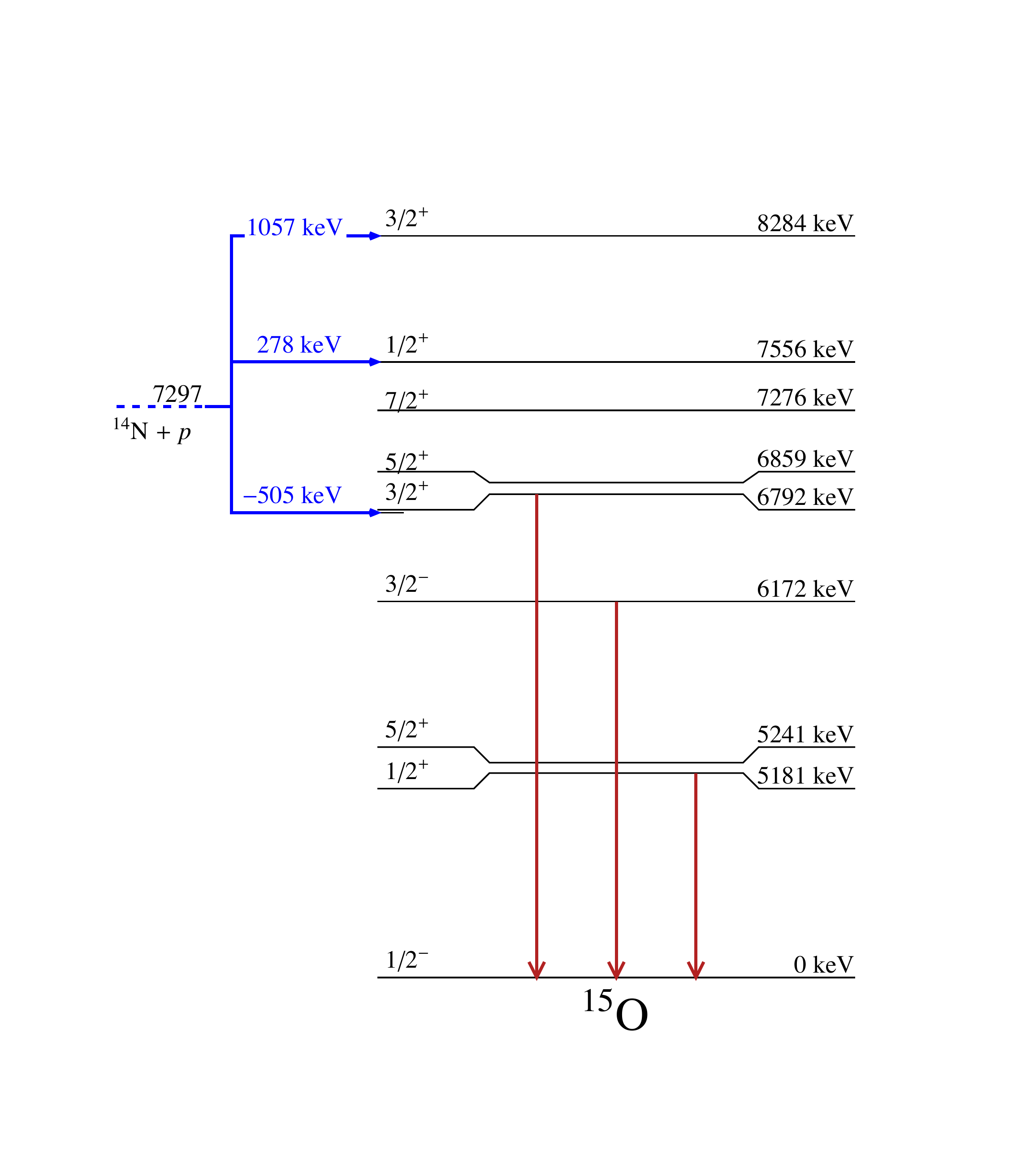}
\caption{Level scheme of the $^{15}$O compound nucleus. The beam (laboratory frame) and resonance (center of mass frame) energies and corresponding excitation energies of the important states are given. In this work, the lifetimes of the states at 6792 keV, 6172 keV, and 5181 keV are reported. These states all decay with 100\% branching to the ground state.}
\label{fig: levelScheme}
\end{figure}

The first comprehensive study of the $^{14}$N$(p,\gamma)^{15}$O reaction was performed by \citet{Schroder1987}, covering the proton energy range from $E_p$ = 0.2 to 3.6~MeV. The total $S$-factor at zero energy, $S$(0), was determined to be 3.20 $\pm$ 0.54~keV~b, where the transitions to the ground state and to the $E_x$ = 6.79~MeV excited state in $^{15}$O were dominant, contributing $S_{g.s.}(0) = 1.55 \pm 0.34$ keV b and $S_{6.79}(0) = 1.41 \pm 0.02$ keV b, respectively. Fig.~\ref{fig: levelScheme} depicts the level structure of the $^{15}$O compound nucleus. However, an $R$-matrix analysis by \citet{Angulo2001} drastically changed the extrapolated $S$-factor for both transitions to $S_{g.s.}(0) = 0.08^{+0.13}_{-0.06}$ keV b and $S_{6.79}(0) = 1.63 \pm 0.17$ keV b. Overall, this reduced the total $S$-factor by a factor of 1.7. A number of more recent measurements at the LUNA underground accelerator  \cite{Imbriani2005, Lemut2006, Bemmerer2006, Marta2008} expanded the reaction data to lower energies, suggesting an even lower $S$-factor, while an independent study by~\citet{Runkle2005} indicated a higher value for the ground state transition. An independent $R$-matrix analysis of the various reaction channels over a wide range of energies by \citet{Azuma2010} and \citet{Li2016} failed to confirm the LUNA collaboration measurements and suggest a higher low energy $S$-factor, in particular for the transition to the ground state. This discrepancy translates into a substantial uncertainty in the prediction of the reaction rate and the associated neutrino flux from the decay of $^{15}$O. % have found similar contributions from these transitions, falling in the range of $S_{g.s.}(0) = 0.15 - 0.49$ keV b. 
The very low energy contributions in the $^{14}$N$(p,\gamma)^{15}$O reaction rate come from the tail contribution of the subthreshold state at $E_{x}$ = 6.79 MeV in $^{15}$O, which is determined by its radiative width, $\Gamma_{\gamma}$.

The 6.79 MeV excited state decays with 100\% probability to the ground state by $\gamma$-ray emission. Therefore, the total width of the state results solely from the radiative width, $\Gamma = \Gamma_{\gamma}$. The width can be determined by measuring the level lifetime, $\tau$, where $\Gamma = \hbar/\tau$. The first precision measurement of this lifetime was performed by \citet{Bertone2001}, using the Doppler Shift Attenuation Method (DSAM) in forward kinematics with targets of N implanted into Ta. They populated the state through resonant capture of protons to the $E_{p}$ = 278 keV resonance of the $^{14}$N$(p,\gamma)^{15}$O reaction and observed the decaying $\gamma$-rays at three different angles. The measured lifetime for the 6.79 MeV state was determined to be $\tau = 1.60^{+0.75}_{-0.72}$ fs, corresponding to a width of $\Gamma = 0.41^{+0.34}_{-0.13}$ eV, where the reported uncertainties correspond to a 90\% confidence limit. The authors also noted that their reported value would have increased by 100\%, to $\tau = 3.2 \pm 1.5$ fs, if they used the density of the TaN compound (as their target was implanted TaN) instead of pure Ta in their calculations. 

\citet{Yamada2004} performed another measurement of the radiative width of this level at RIKEN in 2004,   Coulomb excitation. This measurement of $^{15}$O nuclei was performed by fragmentation of an $^{16}$O beam incident on a $^{9}$Be target and then inelastically scattered off of a thick lead target. The de-excitation $\gamma$-rays were detected with an array of NaI (Tl doped) scintillation detectors. In order to disentangle the 6.79 MeV peak of interest from the 6.86 MeV peak in their spectrum, the authors compared the experimentally obtained spectrum with the results of a Monte Carlo simulation \cite{Yamada2004}. Due to the poor energy resolution of the detector system, the authors ultimately had difficulty separating the peaks and were unable to observe isolated $\gamma$-rays from the 6.79 MeV state. The authors only reported an upper bound on the width, $\Gamma = 0.95^{+0.60}_{-0.95}$ eV. 

\citet{Schurmann2008} published a new measurement of the lifetime of the same state, using the DSAM technique and an improved experimental setup. This measurement used a single high-purity germanium (HPGe) detector on a rotating track and measured the angular distribution of the depopulating $\gamma$-ray at several angles ranging from $40 - 116\degree$, allowing for a check of asymmetries around the $90\degree$ point and reducing systematic uncertainties from the use of different detectors. Similar to the measurement of \citet{Bertone2001}, the authors also used implanted TaN targets and populated the state of interest via the resonance at $E_{p} = 278$ keV in the $^{14}$N$\left( p,\gamma \right) ^{15}$O reaction. They obtained an upper limit of $\tau < 0.77$ fs for the lifetime with 68.3\% confidence. In contrast with the result of \citet{Bertone2001}, however, the authors note only a 2\% change in the reported lifetime by changing the target density to TaN in their analysis. 

\citet{Galinski2014} performed another DSAM measurement at the TRIUMF facility. The major difference between past measurements was that the reaction was performed in inverse kinematics, emulating a measurement from over 50 years ago~\cite{Gill1968}. The $^{15}$O in this experiment was produced via the $^{3}$He$(^{16}$O$,\alpha)^{15}$O reaction with a $^{3}$He-implanted gold-foil. The de-excitation $\gamma$-rays were detected with a clover HPGe detector and filtered in coincidence with the reaction $\alpha$-particles. The authors performed a Bayesian line-shape analysis of the spectra and, using the maximum likelihood method, recommended an upper limit of $\tau < 1.8$ fs with a 68.3\% confidence limit. The authors, however, acknowledge that their result was statistics limited. 

The only other recent measurement of this lifetime comes from a doctoral dissertation by \citet{Michelagnoli2013}. Similar to that of \citet{Galinski2014}, this measurement was done in inverse kinematics, with the $d(^{14}$N$,^{15}$O$)n$ reaction. The $\gamma$-rays from this measurement were detected with the state-of-the-art AGATA Demonstrator array using $\gamma$-ray tracking for precision angular data. A line shape analysis of the data was performed by comparing with the results of a Monte Carlo simulation, also obtaining an upper-bound on the lifetime, $\tau < 1.0$ fs with a 68.3\% confidence limit. 

In aggregate, these results are unsatisfactory due to their disagreement. The DSAM results in forward kinematics had neglected the systematic uncertainties arising from the stopping power of the targets in their ultimate lifetime calculation. The forward kinematics reactions \cite{Bertone2001, Schurmann2008} both used the same type of nitrogen-implanted tantalum-foils for the experiment (saturated to a composition of Ta$_{2}$N$_{3}$), but calculated the lifetimes assuming target densities and stopping powers equal to those of pure tantalum, calculated in SRIM \cite{Ziegler2010}. However, the authors note that accurate densities would have changed the inferred lifetimes of their works by 100\% \cite{Bertone2001} and 2\% \cite{Schurmann2008}, respectively.

This work reports on a new measurement of the the lifetime of the 6.79 MeV excited state in $^{15}$O by populating it with the same $^{14}$N$(p,\gamma)^{15}$O reaction in forward kinematics at proton energies ranging from $E_{p} = 1.0 - 1.6$ MeV using the DSAM technique. Due to the discrepancies between the \citet{Bertone2001} and \citet{Schurmann2008} measurements, discrepancies which occurred despite the use of the same target type, reaction, and method, we have used three different types of nitrogen implanted targets with backings of Mo, Ta, and W to reduce systematic uncertainties. The targets were produced with the 5 MV Sta. Ana accelerator at the University of Notre Dame Nuclear Science Laboratory. We report measurements of the lifetimes of this state as well as the 5.18 MeV and 6.17 MeV states in $^{15}$O, which are in the fs range and thus used to validate the results for the state at 6.79 MeV. 

In this work, the experimental details are first discussed in Sec.~\ref{sec: exp}, covering the target implantation and lifetime measurements. In Sec.~\ref{sec: analysis}, a Monte Carlo simulation is described followed by the calculation of the lifetime via the observed Doppler shifts. In order to demonstrate the effect of the lifetime measurement and its uncertainty on the $^{14}$N$(p,\gamma)^{15}$O reaction, preliminary $R$-matrix calculations are described in Sec.~\ref{sec: rmatrix}. Summary and conclusions are presented in Sec.~\ref{sec: summary}.

\section{Experiment }
\label{sec: exp}

\subsection{Implanted targets}
\label{sec: targets}

Isotopically pure targets were made by implantation using the 5 MV Sta. Ana accelerator at the University of Notre Dame's Nuclear Science Laboratory (NSL). 350 keV $^{14}$N$^{1+}$ ions were implanted into Ta, W, and Mo backings of 0.5 mm thickness. %The implantation was done at 350 keV for accelerator stability reasons.  
A similar implantation technique was used in another study \cite{Seuthe1987} that showed a deeper distribution of nitrogen in the target. The Ta was chosen to compare with the earlier measurements of \cite{Bertone2001, Schurmann2008}, which used Ta$_{2}$N$_{3}$, while the W and Mo were used to identify and highlight any systematic differences arising from the choice of backing material. This also provided a method of disentangling the effects of the high beam currents used in the measurement.   
Before the implantation, the target backings were cut and cleaned using ethanol, acetone, and with an oxygen plasma. Throughout the implantation, the beam was rastered over the backing surface in order to produce a uniform implantation. Additionally, in order to reduce the carbon buildup on the target face during the implantation, a liquid-nitrogen cooled copper cold trap was employed. Due to the beam intensities used, the target backings were water cooled. 

The nitrogen content in the targets was assessed by scanning the 1057 keV resonance in the $^{14}$N$(p,\gamma)^{15}$O reaction and comparing the yield with that of a target previously used by \citet{Li2016}. The Ta, Mo, and W targets were found to have nitrogen contents of 21.3 $\pm$ 1.3, 13.1 $\pm$ 0.7, and 13.4 $\pm$ 0.5 $\times 10^{17}$ atoms / cm$^{2}$. Earlier targets were produced to saturation of the nitrogen in interstitial lattice sites, resulting in targets of 60\% nitrogen. These targets, however, were produced to 36, 26, and 22 percent, respectively for the Ta, Mo, and W targets. As the backing materials saturate with nitrogen to different compounds, a lower nitrogen content was used to allow for improved consistency between the different targets. 

\subsection{Lifetime Measurement Method}
\label{sec: setup}

The excited 5.18 MeV, 6.17 MeV, and 6.79 MeV excited states in $^{15}$O were populated with the $^{14}$N$(p,\gamma)^{15}$O reaction, where protons of $E_{p}$ = 1020 keV and $E_{p}$ = 1570 keV were delivered to the targets. At both energies, the reaction proceeds primarily through the direct-capture mechanism and these energies were chosen to provide the best signal-to-background discrimination for the states of interest. The energy spread of the beam was approximately 1 keV and the current on target throughout the experiment ranged from 20 - 50 $\mu$A, dependent on accelerator stability. Targets were mounted on a $45\degree$ target holder, relative to the beam axis, and due to the high beam currents, the backings were constantly cooled with recirculating de-ionized water. A copper cold finger, biased to $-400$~V and cooled with a liquid nitrogen reservoir, was utilized to limit carbon build-up and suppress secondary electrons throughout the measurements. To keep the beam focused and centered through the runs, some of which lasted several hours, pairs of upstream slits were utilized. These prevented the beam from drifting to the edges of the implanted area on the targets where the nitrogen content was lower.

\begin{figure}
\includegraphics[width=1.0\columnwidth]{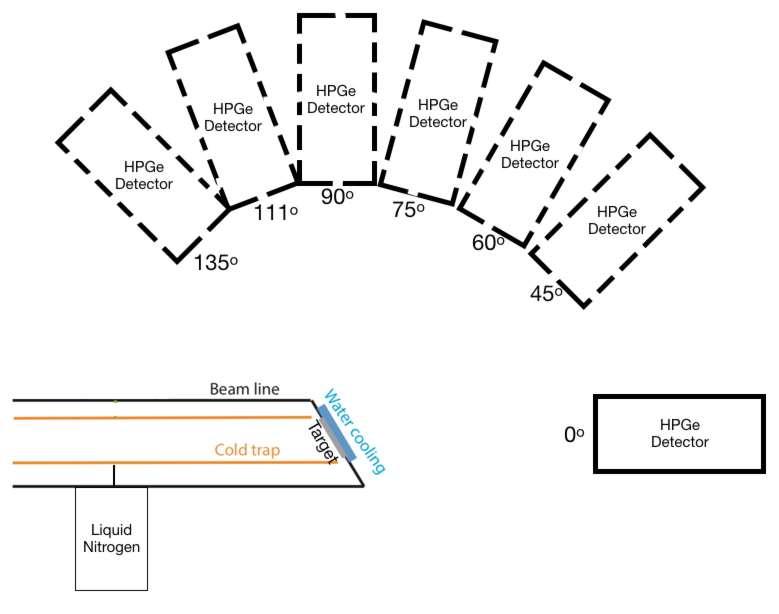}
\caption{Schematic of the experimental setup. It is important to note that the same HPGe detector was set on a rotating table and set to each of the angles, not seven different detectors. }
\label{fig: setup}
\end{figure}

\begin{figure}
\includegraphics[width=1.0\columnwidth]{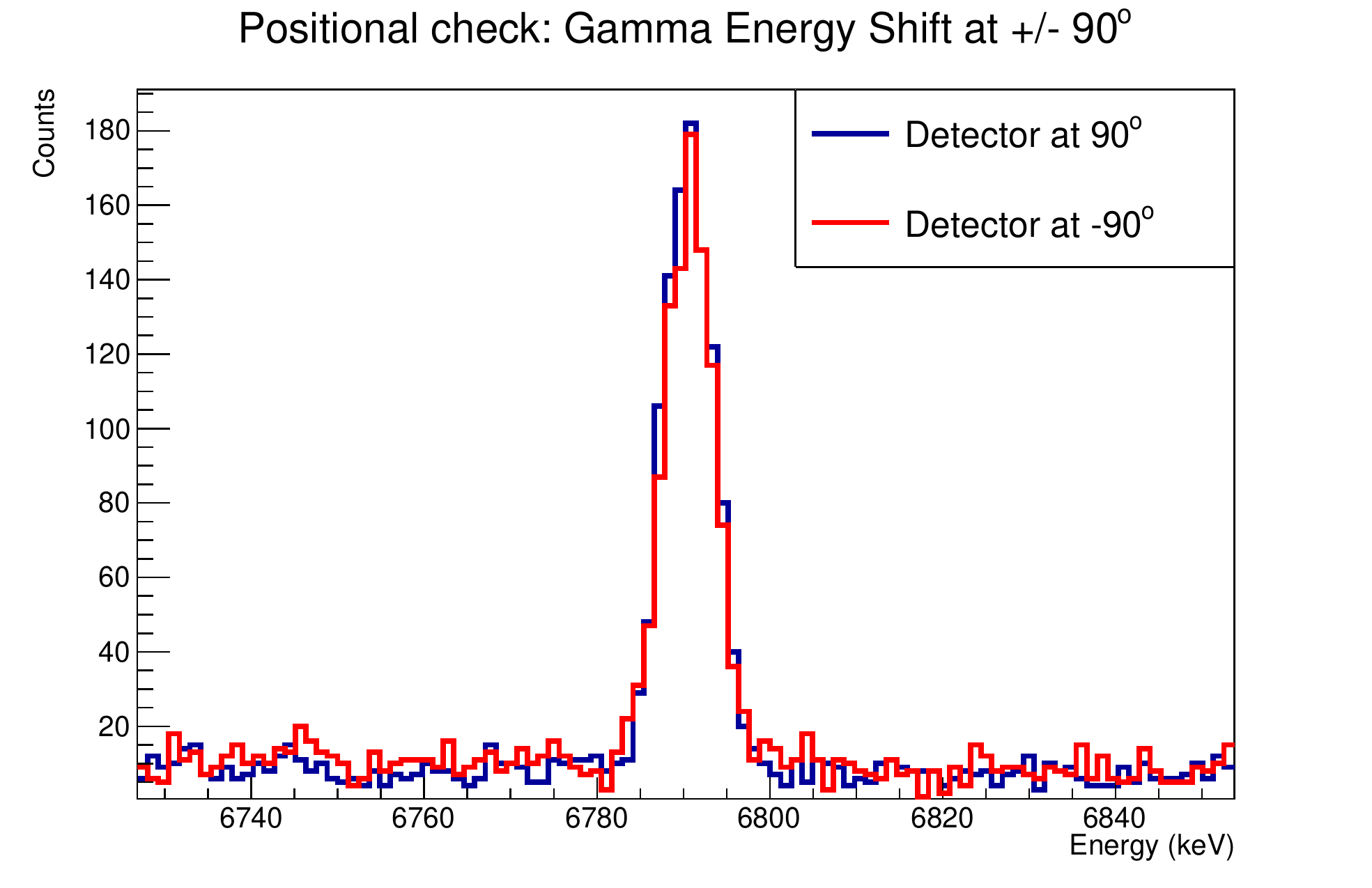}
\caption{Measured energy spectrum of the 6.79 MeV transition at $\pm$ 90$\degree$. At these two angles, the energy of the $\gamma$ ray is not Doppler shifted. Therefore, by comparing the position of the peaks at these angles, we confirm accurate position of the detector relative to the target and centering of the beam on target.}
\label{fig: 90compare}
\end{figure}

The de-exciting $\gamma$ rays were observed with one of the GEORGINA detectors, a coaxial n-type HPGe detector, of 109\% relative efficiency. 1.5 mm lead shielding was placed in front of the crystal's face to attenuate the low-energy x rays entering the detector, reducing the count rate from the low energy background. The lifetimes were measured by the DSAM method. A single detector was used to measure the depopulating $\gamma$-rays at different angles of $0\degree, 45\degree, 60\degree, 75\degree, 90\degree, 111\degree, 135\degree$, and $-90\degree$ relative to the beam direction. The detector was placed 25.4 cm away from the target on a rotating table with angular precision such that the detector's center could be placed within $\pm$ 0.5$\degree$ of the intended angle. Fig.~\ref{fig: setup} shows a schematic of the experimental setup. Data at $-90\degree$ were taken and compared to those at $90\degree$ to prove that the beam was centered on target, since $\gamma$ rays detected at these corresponding angles are unshifted. These two spectra are included in Fig.~\ref{fig: 90compare}. Additionally, the data at $90\degree$ were used to calibrate the spectra in the detector.

In order to prevent any peak shifting from smearing the peaks at a given angle, the data were saved hourly to track the stability of the detector. This allowed the monitoring of the detector for gain shifts throughout the experiment. Additionally, a CeBr detector was fixed at a backward angle of $-135\degree$ in order to provide a consistency check throughout the experiment and monitor for any drifting. By monitoring background lines, we found no signs of instability of the detector or electronics. The data from both detectors were collected with a Mesytec MDPP-16 module, which amplified and digitized the data. 

\section{Analysis and results}
\label{sec: analysis}

\subsection{Monte Carlo simulation}

\begin{figure}
\includegraphics[width=1.0\columnwidth]{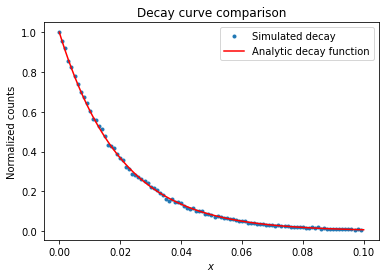}
\caption{An example of the Monte Carlo method for a simulated decay curve. For 100,000 decays with $\lambda$ = 50 (chosen simply for demonstration), the normalized decay curve agrees with the exponential decay law.}
\label{fig: mcDemonstration}
\end{figure}

For this analysis, a DSAM program was written to simulate the expected Doppler shift. The approach utilized Monte Carlo sampling to simulate the radioactive decay of a nucleus with a given lifetime according to the normal exponential decay law:
\begin{equation}
P(t) = e^{-t/\tau}.
\label{eqn: decay}
\end{equation} 
The generator function for exponential decay, takes the random number, $r$, within the range [0, 1], and propagates it to the simulated decay time through 
\begin{equation}
t = \dfrac{1}{\lambda} \ln \left( \dfrac{1}{1-r} \right),
\label{eqn: gen}
\end{equation}
where $\lambda = 1/\tau$ is the width of the state. To demonstrate this, Fig.~\ref{fig: mcDemonstration} provides an example of the simulated decay probability as a function of time for radioactive nuclei following the decay law of Eq.~(\ref{eqn: decay}).

These simulated lifetimes (with the target backings etc.) were used in conjunction with the output from the SRIM stopping power software \cite{Ziegler2010}. Targets were generated in the software and then used to simulate ion recoil tracks within the target after the reaction. The initial energy of the recoil was randomly assigned to a reasonable range based on the energy loss in the target and all recoils were assumed to have an initial velocity parallel to the incident beam axis. For 10,000 different ions, the entire travel path of the recoil through the target was simulated and saved. The energy of the recoiling nuclei, in the case of this experiment, was not high enough to warrant relativistic corrections, like the method of \cite{Galinski2014}, so the SRIM output was sufficient. 

Based on the SRIM simulation, further information about each individual ion's travel was calculated. Utilizing the position of each interaction, the particle's trajectory was calculated between each position. The recorded energy at each interaction was used to calculate the ion's speed in the step. Finally, these data allow for the time of the interaction to be calculated (in femtoseconds). This provided a complete picture of any single recoiling ion's motion in the target. Additionally, in \citet{Ziegler2010}, the SRIM software quotes stopping power uncertainties of 5\%. The code took this uncertainty into account and propagated it through to an uncertainty in the initial energy of the recoil ion. 

Next, the program used Eq.~(\ref{eqn: gen}) to randomly simulate a decay time to the decaying nucleus (for a specified lifetime) and randomly assigned this decay time to one of the SRIM tracks. Then, the code interpolated between interaction points to find the position and velocity of the decaying nucleus at the generated instant of decay. With this information, the Doppler shifted de-exciting $\gamma$ energy, $E_{\gamma}$ was calculated by
\begin{equation}
E_{\gamma} = E_{\gamma}^{o} \left( 1 + R \beta F(\tau) \cos (\theta ) \right),
\end{equation}
where $E_{\gamma}^{o}$ was the unshifted $\gamma$ ray energy, $R$ represents a correction to account for the detector size and resolution (obtained from experimental spectra), $\beta = v / c$ was the recoil's velocity relative to the speed of light, $F(\tau)$ was the attenuation factor (encapsulating the relationship between the lifetime of the decaying state and the medium through which the nucleus moves), and $\theta$ was the angle at which the $\gamma$ was observed. This process was repeated 50,000 times for each combination of detection angle, lifetime, and target backing material in order to build up a complete Monte Carlo simulation of the experiment. 

\begin{figure}
\includegraphics[width=1.0\columnwidth]{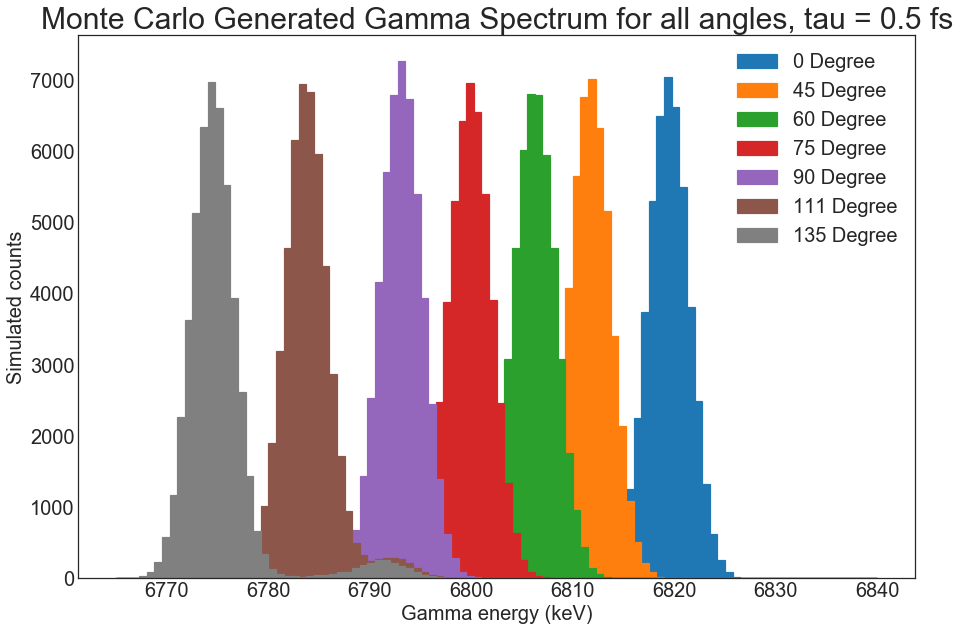}
\caption{Simulated energy deposition histogram for the 6.79 MeV state de-excitation, at all detection angles, $\tau = 0.5$ fs, and the Mo target backing. This figure clearly illustrates the simulated Doppler shifting of the $\gamma$-rays' energy with angle, the relationship upon which the entire DSAM was predicated.   }
\label{fig: exampleMC}
\end{figure}

\begin{figure}
\includegraphics[width=1.0\columnwidth]{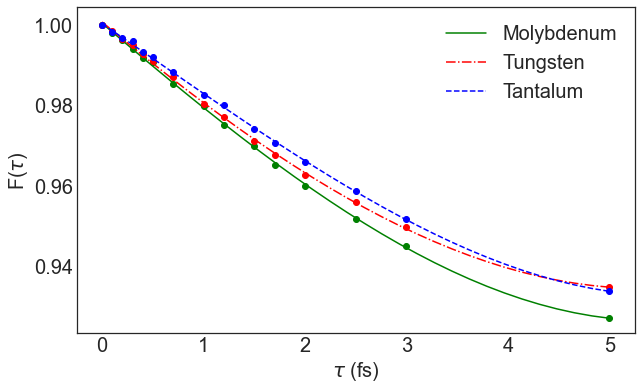}
\caption{The simulated attenuation factors, $F(\tau)$, for the various experimental backings, as a function of the state's lifetime, $\tau$. This was used to calculate the measured lifetime from the experimental attenuation factors. The simulations were also carried out to lifetimes well beyond what was expected from literature values to ensure that the relationship was robust around the lifetimes of interest. }
\label{fig: attFactors}
\end{figure}

An example of a complete simulation for the 6.79 MeV state de-excitation at all angles, $\tau = 0.5$ fs, and the Mo target backing is shown in Fig.~\ref{fig: exampleMC}, demonstrating the Doppler shift of the peak energy with detection angle and demonstrating correct behavior with an energy boost at forward angles and a reduction in energy at backward angles. Using the relationship between the centroid of the peak and angle, as generated by these simulations, the attenuation factor, $F(\tau)$, was determined as a function of the nuclear lifetime, $\tau$, for each individual target. This relationship is plotted in Fig.~\ref{fig: attFactors}. This was ultimately used to determine the experimental lifetime by comparing the measured attenuation factors to those of the simulated curves to extract the corresponding lifetime and uncertainty.

A significant difference between this procedure and other, previous methods for calculating the lifetimes was that they only used numerical calculations. This method is only a discrete calculation at values chosen across the relevant lifetime landscape. Therefore, these attenuation factors require an external fit to provide the final relationship. To minimize any artificial bias from the choice of simulated lifetimes, they were chosen to range beyond expected values to ensure that the relationship between the lifetime and attenuation factor is robust in the relevant lifetime range.

\subsection{Doppler shifts and lifetimes}

\begin{figure}
\includegraphics[width=1.0\columnwidth]{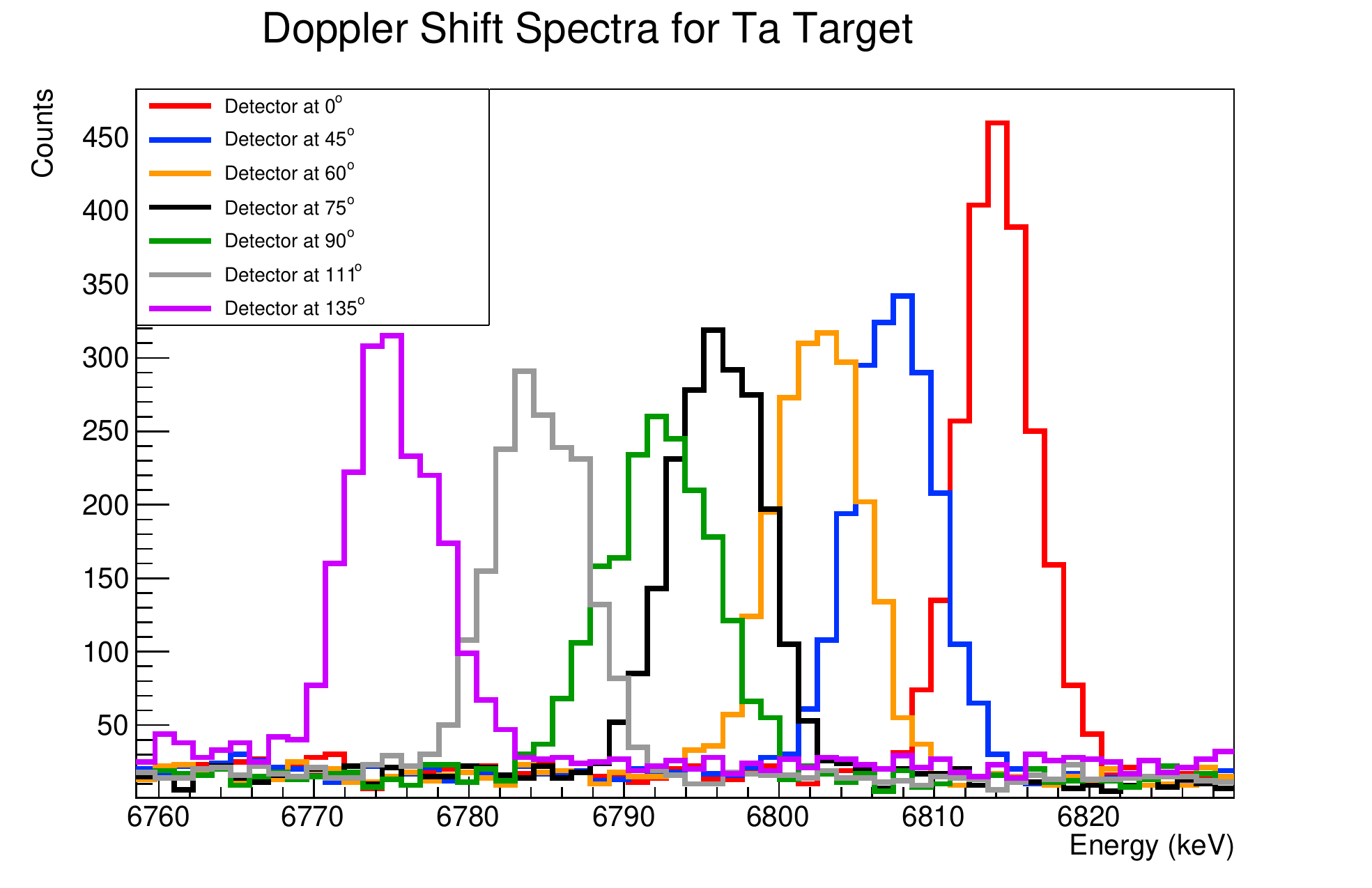}
\caption{Doppler shifting spectra for the 6.79 MeV secondary transition with detection angle for the Ta target, with all angles ($0\degree, 45\degree, 60\degree, 75\degree, 90\degree, 111\degree$ and $135\degree$) plotted together.  }
\label{fig: gammaShift}
\end{figure}

\begin{figure}
\includegraphics[width=1.0\columnwidth]{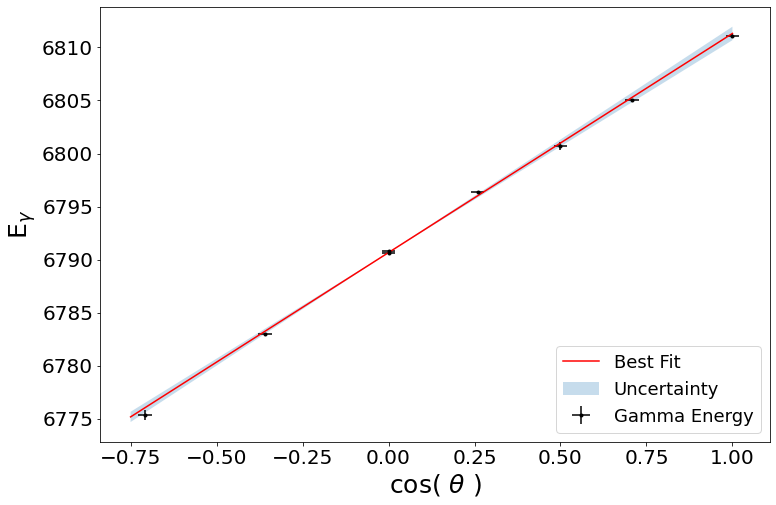}
\caption{Doppler shifted $\gamma$-ray energy plotted against $\cos(\theta)$ to determine $F(\tau)$, the attenuation factor. The red line represents the orthogonal distance regression best fit and the band represents a two-sigma uncertainty of the fit, incorporating the uncertainty in both the slope and intercept.  }
\label{fig: doppler}
\end{figure}

In the measurement at each of the seven angles ($0\degree, 45\degree, 60\degree, 75\degree, 90\degree, 111\degree$ and $135\degree$), the centroid energy of each of the three secondary transition peaks at 5.18 MeV, 6.17 MeV, and 6.79 MeV were determined. Example spectra of the peak Doppler shifting is shown in Fig.~\ref{fig: gammaShift}. From this, the relationship between the cosine of the measurement angle, $\cos(\theta)$, and the observed energies, $E_{\gamma}$, was determined. With uncertainties in both the angle and the centroid energy, the data were fit with orthogonal distance regression, instead of the traditional least-squares method, because it takes into account the uncertainties in both dependent and independent variables, leading to higher fidelity fits. The relationship between shifted $\gamma$-ray energy and $\cos(\theta)$ is shown in Fig.~\ref{fig: doppler}. These figures are representative of the other two targets as well.

The experimentally determined attenuation factors and corresponding lifetimes are shown in Table~\ref{table: attFacs}. Additionally, the weighted average of the lifetimes has also been calculated. This analysis leads to lifetimes for each of the 5.18 MeV, 6.17 MeV, and 6.79 MeV excited states in $^{15}$O. Table~\ref{table: attFacs} results show no systematic differences with the lifetimes extracted from the various target materials. This allowed the exclusion of backing materials as sources of uncertainties in comparison to earlier measurements.

\begin{table}

\caption{Experimental $F(\tau)$ and corresponding lifetime values obtained in the present work along with a weighted average of the lifetime for the various targets. Note that since the attenuation factor is dependent on the backing material, only the average of the lifetimes may be used. }
\begin{center}
\begin{ruledtabular}
\begin{tabular}{lllll}
                   & Mo                   & Ta                  & W                   & $\tau_{\text{average}}$ \\ \hline 
$F(\tau)_{5.18}$   & $9.040 \pm 0.013$     & $0.911 \pm 0.016$   & $9.120 \pm 0.015$    &                  \\
$\tau_{5.18}$ (fs) & $﻿7.1^{+4.8}_{-2.3}$ & $7.1 \pm 5.5$       & $8.0 \pm 6.7$       & $7.5 \pm 3.0$    \\ \hline
$F(\tau)_{6.17}$   & $0.992 \pm 0.014$    & $0.976 \pm 0.017$   & $0.988 \pm 0.016$   &                  \\
$\tau_{6.17}$ (fs) & $0.4^{+0.7}_{-0.4}$  & $1.4 \pm 1.0$       & $0.6^{+0.9}_{-0.6}$ & $0.7 \pm 0.5$    \\ \hline
$F(\tau)_{6.79}$   & $0.995 \pm 0.019$    & $ 0.983 \pm 0.019$  & $0.978 \pm 0.015$   &                  \\
$\tau_{6.79}$ (fs) & $0.2^{+0.7}_{-0.2}$  & $0.7^{+0.9}_{-0.7}$ & $0.9 \pm 0.6$       & $0.6 \pm 0.4$    \\  
\end{tabular}
\end{ruledtabular}
\label{table: attFacs}
\end{center}
\end{table}

\begin{figure}[ht]
\includegraphics[width=1.0\columnwidth]{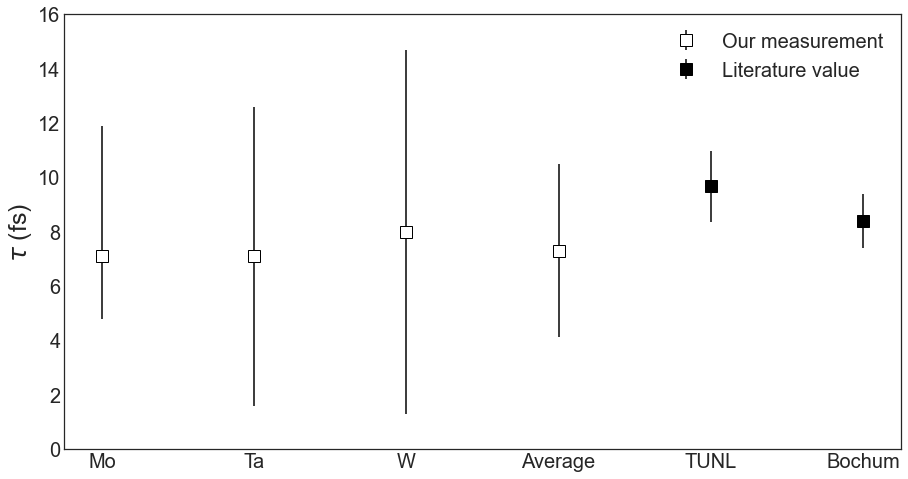}
\caption{Lifetimes of the 5.18 MeV state in $^{15}$O obtained in this work compared with previous measurements. The measurement labelled TUNL corresponds to that of \citet{Bertone2001}, while Bochum to \citet{Schurmann2008}. }
\label{fig: lifetimes518}
\end{figure}

\begin{figure}[ht]
\includegraphics[width=1.0\columnwidth]{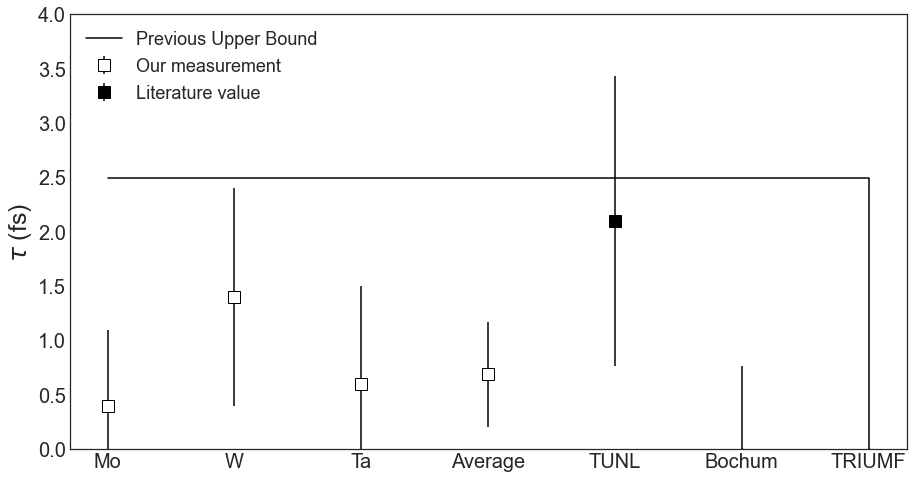}
\caption{Lifetimes of the 6.17 MeV state in $^{15}$O obtained in this work compared with previous measurements. The measurement labelled TUNL corresponds to that of \citet{Bertone2001}, Bochum to \citet{Schurmann2008}, and TRIUMF to \citet{Galinski2014}. }
\label{fig: lifetimes617}
\end{figure}

\begin{figure}[ht]
\includegraphics[width=1.0\columnwidth]{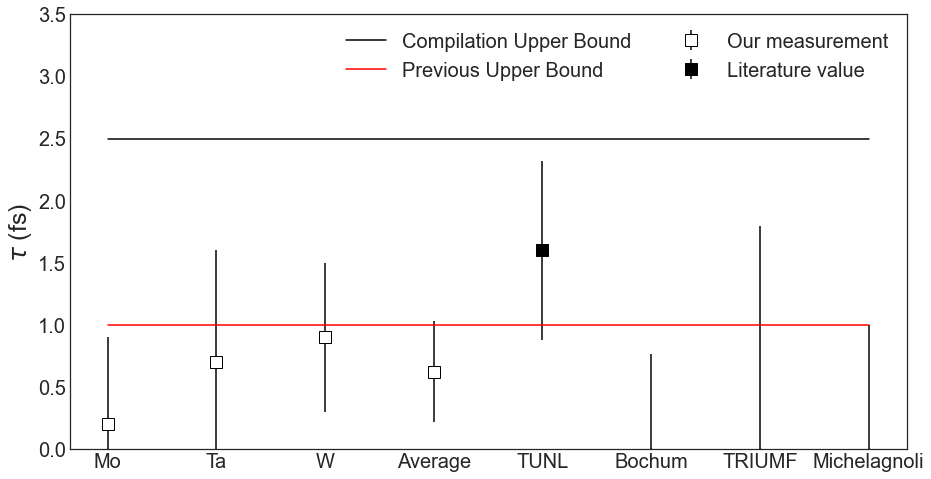}
\caption{Lifetimes of the 6.79 MeV state in $^{15}$O obtained in this work compared with previous measurements. The measurement labelled TUNL corresponds to that of \citet{Bertone2001}, Bochum to \citet{Schurmann2008}, TRIUMF to \citet{Galinski2014} and Michelagnoli to \citet{Michelagnoli2013}, which has not been peer reviewed. }
\label{fig: lifetimes679}
\end{figure}

The experimentally determined lifetimes are shown in comparison with previous measurements in Table~\ref{table: lifetimes}. The mean lifetime of the 5.18 MeV state agrees well with previous measurements, but has a much larger uncertainty. This large uncertainty arises in part from the double escape peak at $E_{\gamma} = 5150$ keV (from the $E_{\gamma} = 6172$ keV full energy peak) that complicates the background determination (and therefore the peak centroids for the Doppler Shift). For the lifetime of the 6.17 MeV and 6.79 MeV states, we find good agreement with the results reported in \citet{Schurmann2008} and \citet{Galinski2014}. For each of the states, our recommended lifetimes are taken to be the weighted average of the measurements taken from each of the targets. These results are depicted in Figs.~\ref{fig: lifetimes518}, \ref{fig: lifetimes617}, and \ref{fig: lifetimes679}.

\begin{table*}

\caption{Experimentally determined lifetime values (in fs) obtained in this work (as the weighted average of the different targets) and their comparison with literature values.}
\begin{center}
\begin{ruledtabular}
\begin{tabular}{lllll}
$E_{x}$ (keV) & Present       & Ref. \cite{Bertone2001}  & Ref. \cite{Schurmann2008} & Ref. \cite{Galinski2014} \\
5181          & $7.5 \pm 3.0$ & 9.67$^{+1.34}_{-1.24}$ & 8.40$\pm$1.00            & -                       \\
6172          & $0.7 \pm 0.5$ & 2.10$^{+1.33}_{-1.32}$  & $< 0.77$                 & $< 2.5$                 \\
6793          & $0.6 \pm 0.4$ & 1.60$^{+0.75}_{-0.72}$  & $< 0.77$                 & $< 1.8$                
\end{tabular}
\end{ruledtabular}
\label{table: lifetimes}
\end{center}
\end{table*}

\section{$R$-Matrix Fit}
\label{sec: rmatrix}

\begin{figure}[h!]
\includegraphics[width=1.0\columnwidth]{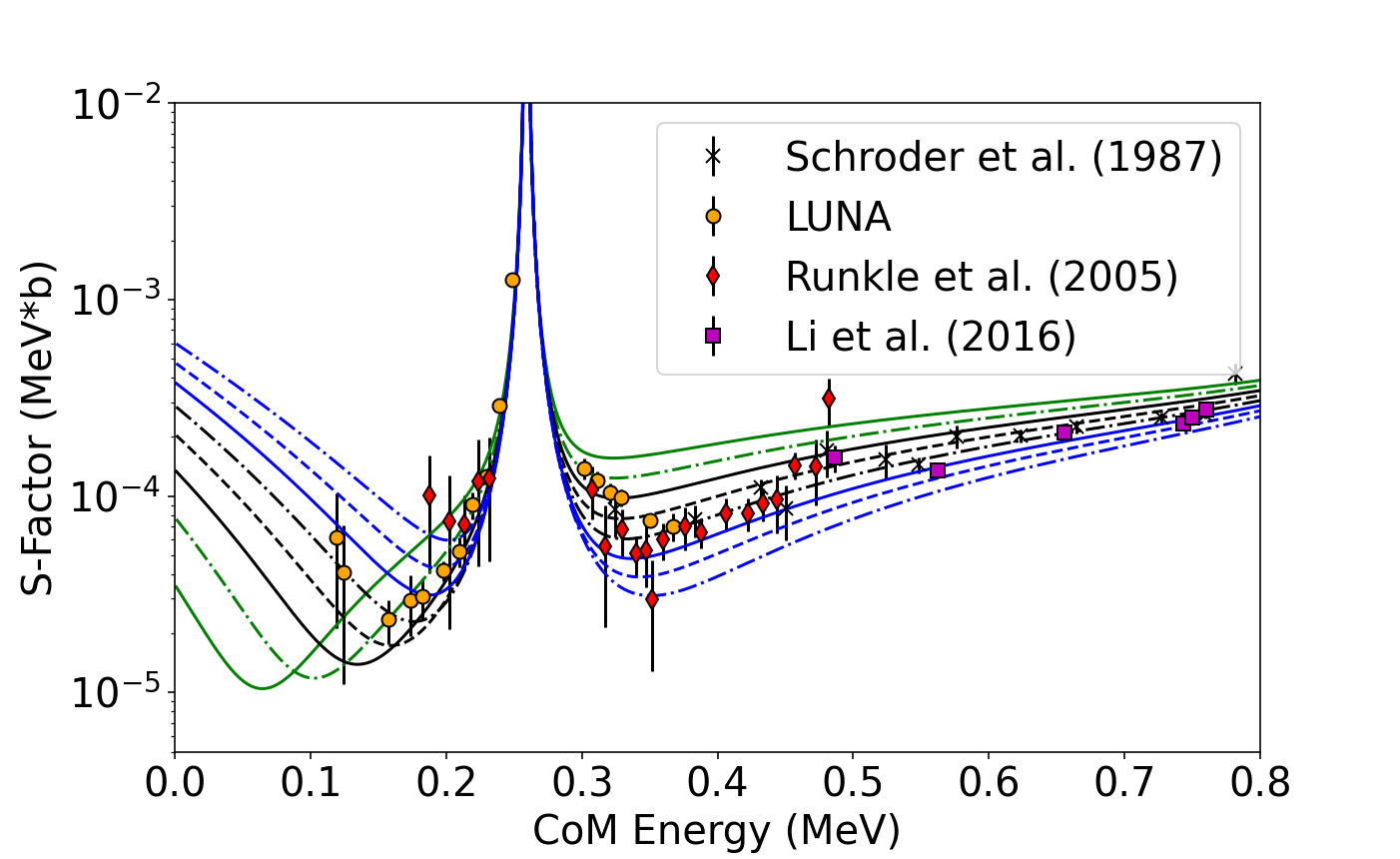}
\caption{$R$-Matrix fits exploring the uncertainty of our lifetime measurements to the low energy extrapolation. The width of the 6.79 MeV excited state in $^{15}$O is fixed during each fit and changed in each subsequent iteration to another value within our uncertainty range. This clearly shows that even though our lifetime result provides the most stringent limitation on the lifetime of this state, it still has an outsized effect on the low energy behavior of this reaction. The Schr{\"{o}}der data are from \cite{Schroder1987}, while the LUNA data represents the measurements \cite{Formicola2004, Imbriani2005, Marta2008, Marta2011}, the Runkle data are from \cite{Runkle2005}, and the Li data are from \cite{Li2016}. Of these, the Li data are differential and were treated as such in the fitting but scaled up by 4$\pi$ for plotting purposes.}
\label{fig: rmatrixRange}
\end{figure}

\begin{figure}[h!]
\includegraphics[width=1.0\columnwidth]{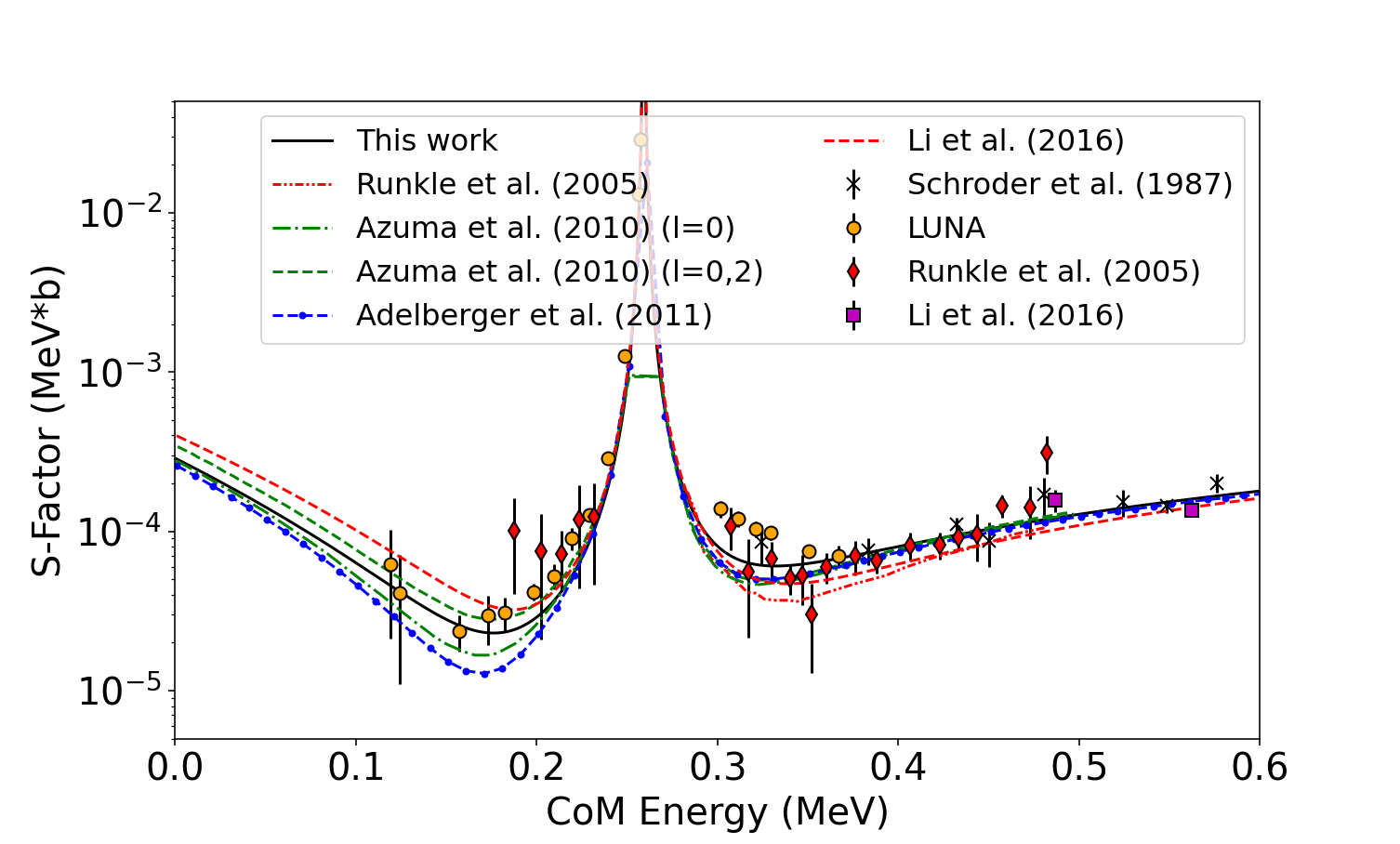}
\caption{$R$-Matrix fits comparing our best fit to those performed in previous works. Our fit used a lifetime value for the 6.79 MeV excited state in $^{15}$O within our measured range range, $\Gamma=2.75$ where the fits to the data provide good agreement with data above and below the 278 keV resonance. This plot is limited to the low energy region. The Schr{\"{o}}der data are from \cite{Schroder1987}, while the LUNA data represents the LUNA measurements \cite{Formicola2004, Imbriani2005, Marta2008, Marta2011}, the Runkle data are from \cite{Runkle2005}, and the Li data are from \cite{Li2016}. Of these, the Li data are differential and were treated as such in the fitting but scaled up by 4$\pi$ for plotting purposes. The fits from previous works come from Refs.~\cite{Runkle2005, Azuma2010, Adelberger2011, Li2016}.}
\label{fig: rmatrixClose}
\end{figure}

To examine the impact of these new lifetime measurements, the data were incorporated into a set of limited $R$-Matrix fits using the \texttt{AZURE2} code \cite{Azuma2010}. These represent fits to a selection of data to study the influence of this lifetime measurement in particular; a full $R$-Matrix fit incorporating these results and new low-energy cross-section results will be presented in a forthcoming paper. In the fits presented here, shown in Figs.~\ref{fig: rmatrixRange} and \ref{fig: rmatrixClose}, most of the information about the levels was taken from  \citet{Ajzenberg-Selove1991} or \citet{Daigle2016} where updated. The width of the 6.79 MeV excited state in $^{15}$O, corresponding to values within the range of our lifetime measurement, was fixed at a selection of different values. This collection of fits, therefore, serves primarily as an illustration of the ways in which these new lifetime measurements impact the low energy extrapolations of the cross section.

A channel radius of 5.5 fm was adopted for this work, which matches the analyses done by \cite{Adelberger2011}, \cite{Li2016}, and \cite{Wagner2018}. Information about the levels and their parameters as used in \texttt{AZURE2} are contained in Table~\ref{table: fitParams}. 

The cross-section data utilized in the fitting routine were from measurements at LUNA \cite{Formicola2004, Imbriani2005, Marta2008, Marta2011}, TUNL \cite{Runkle2005}, Bochum \cite{Schroder1987}, and the University of Notre Dame \cite{Li2016}. All of these data sets were left without scaling during the fits. The Bochum data from \citet{Schroder1987} were corrected as detailed in SFII \cite{Adelberger2011}. Additionally, the data from \citet{Li2016} are a differential cross section taken at 45$\degree$ and are treated as such in the fits. This dataset was scaled by a factor of 4$\pi$ in the plotting only to compare to the angle integrated data.

\begin{table*}
\caption{Levels used in the \textit{R}-matrix fits. Bold values indicate parameters which were allowed to vary during the fit. The signs on the partial widths and ANCs indicates the relative interferences. The dividing line demarcates the proton separation energy at $E_x$ = 7.2968(5) MeV \cite{Ajzenberg-Selove1991}. Levels where all parameters are fixed are not shown in this table for brevity but were included in the fits. a) Indicates the partial width of the 6.79 MeV state, measured in this experiment to between $\Gamma$ = 0.66 - 3.29 eV. For each individual fit, this width was fixed. However, between each fit, this width was varied to different values within our range to explore how the uncertainty in this measurement affects the low energy extrapolation. These different fits are shown in Fig.\ \ref{fig: rmatrixRange} and are otherwise identical. }
\begin{center}
\begin{ruledtabular}
\begin{tabular}{c  c  c  c  c  c  c}
$E_x$ (Ref.~\cite{Ajzenberg-Selove1991}) &   $E_x$ (fit) & $J^\pi$ & Channel & l & s & ANC (fm$^{-1/2}$) / Partial Width (eV) \\ 
   \hline
0.0 & 0.0	& 1/2$^-$ &	$^{14}$N+p &	1&	1/2&	{0.23}\\
	&	&	    &    $^{14}$N+p &	1&	3/2&	{7.4} \\
%5.183(1) & \textbf{5.183}&	1/2$^+$&	$^{14}$N+p&	0&	1/2&	\textbf{0.33}\\
%	&	&		    &$^{15}$O+$\gamma_{0.00}$  &	E1&	1/2&	\textbf{0.0784}\\
%5.2409(3) & \textbf{5.2409}&	5/2$^+$&	$^{14}$N+p &	2&	1/2&	\textbf{0.23}\\
%                &	&		  & $^{14}$N+p &	2&	3/2&	\textbf{0.24}\\
%	 			&	&	 &$^{15}$O+$\gamma_{0.00}$	&M2&	1/2&	\textbf{0.0002}\\
%6.1763(17) & \textbf{6.1763}&	3/2$^-$& $^{14}$N+p &	1&	1/2&	\textbf{0.47}\\
%				&	&	& $^{14}$N+p	&1	&3/2	&\textbf{0.53}\\
%				&	&	&$^{15}$O+$\gamma_{0.00}$	&M1	&1/2&	\textbf{0.865}\\
6.7931(17) & {6.7931}&	3/2$^+$ & $^{14}$N+p &	0&	3/2&	{4.75}\\
	&	&	     &  $^{15}$O+$\gamma_{0.00}$	&  E1  &	1/2&	\textbf{2.50$^{\text{a}}$}\\
%	& &       &  $^{15}$O+$\gamma_{6.17}$	&  E1  &	3/2&{-0.002}\\
%6.8594(9) & \textbf{6.8594}&	5/2$^+$&	$^{14}$N+p&	2&	1/2&\textbf{0.39}\\
%	&		&    &     $^{14}$N+p&	2&	3/2&	\textbf{0.42}\\
%	&		&    &     $^{15}$O+$\gamma_{5.24}$&	M1&	5/2&	\textbf{0.04}\\
%7.2759(6) & \textbf{7.2759}&	7/2$^+$&	$^{14}$N+p &	2&	3/2&	\textbf{1541}\\
%	&		&    &     $^{15}$O+$\gamma_{5.24}$&	M1&	5/2&	\textbf{0.00099}\\
\hline
%7.5565(4) & 7.5563	&	1/2$^+$	&	$^{14}$N+p	&	0	&	1/2	&	\textbf{1.0$\times$10$^3$}	\\
%	&	&	&	$^{15}$O+$\gamma_{0.00}$	&	E1	&	1/2	&	\textbf{0.61$\times$10$^{-3}$}\\
%	&	&	&	$^{15}$O+$\gamma_{6.79}$	&	M1	&	3/2	&	\textbf{8.22$\times$10$^{-3}$}\\
%	&	&	&	$^{15}$O+$\gamma_{5.18}$	&	M1	&	1/2	&	\textbf{0.006}\\
%	&	&	&	$^{15}$O+$\gamma_{6.17}$	&	E1	&	3/2	&	\textbf{0.0254}\\
8.2840(5)& \textbf{8.2848}&	3/2$^+$	&	$^{14}$N+p	&	2	&	1/2	&	{-92.2}\\
	&	&	&	$^{14}$N+p	&	0	&	3/2	&	\textbf{4.013$\times$10$^3$}\\
	&	&	&	$^{14}$N+p	&	2	&	3/2	&	{-509}\\
	&	&	&	$^{15}$O+$\gamma_{0.00}$	&	E1	&	1/2	&	\textbf{0.244}\\
%	&	&	&	$^{15}$O+$\gamma_{5.18}$	&	M1	&	1/2	&	{0.01}\\
%	&	&	&	$^{15}$O+$\gamma_{5.24}$	&	M1	&	5/2	&	{0.2}\\
%	&	&	&	$^{15}$O+$\gamma_{6.17}$	&	E1	&	3/2	&	{-4$\times$10$^{-3}$}\\
%	&	&	&	$^{15}$O+$\gamma_{6.86}$	&	M1	&	5/2	&	{0.01}\\
%8.743(6) & \textbf{8.7502}&	1/2$^+$	&	$^{14}$N+p	&	0	&	1/2	&	\textbf{35.726$\times$10$^3$}\\
%	&	&	&	$^{15}$O+$\gamma_{5.18}$	&	M1	&	1/2	&	\textbf{-0.2}\\
%	&	&	&	$^{15}$O+$\gamma_{6.17}$	&	E1	&	3/2	&	\textbf{0.0827}\\
%8.922(2) & \textbf{8.9219}&	5/2$^+$	&	$^{14}$N+p	&	2	&	3/2	&	\textbf{3.8$\times$10$^3$}\\
%	&	&	&	$^{15}$O+$\gamma_{6.79}$	&	M1	&	3/2	&	\textbf{0.003}\\
8.9821(17) & {8.98}&	5/2$^-$	&	$^{14}$N+p	&	1	&	3/2	&	\textbf{-5.872$\times$10$^3$}\\
	&	&	&	$^{15}$O+$\gamma_{0.00}$	&	E2	&	1/2	&	\textbf{-0.303}\\
	&	&	&	$^{15}$O+$\gamma_{6.79}$	&	E1	&	3/2	&	{-0.001}\\
9.484(8) & \textbf{9.488}&	3/2$^+$	&	$^{14}$N+p	&	2	&	1/2	&	{77.69$\times$10$^3$}	\\
	&	&	&	$^{14}$N+p	&	0	&	3/2	&	\textbf{126.685$\times$10$^3$}	\\
	&	&	&	$^{14}$N+p	&	2	&	3/2	&	{-7.822$\times$10$^3$}\\
	&	&	&	$^{15}$O+$\gamma_{0.00}$	&	E1	&	1/2	&	\textbf{6.92}\\
%	&	&	&	$^{15}$O+$\gamma_{6.86}$	&	M1	&	5/2	&	{0.2}\\
9.488(3) & {9.4905}&	5/2$^-$	&	$^{14}$N+p	&	3	&	1/2	&	{0.979$\times$10$^3$}\\
	&	&	&	$^{14}$N+p	&	1	&	3/2	&	{-6.576$\times$10$^3$}\\
	&	&	&	$^{14}$N+p	&	3	&	3/2	&	{-0.985$\times$10$^3$}\\
	&	&	&	$^{15}$O+$\gamma_{0.00}$	&	E2	&	1/2	&	\textbf{-0.307}\\
	&	&	&	$^{15}$O+$\gamma_{6.79}$	&	E1	&	3/2	&	{-0.0123}\\
9.609(2) & {9.6075}&	3/2$^-$	&	$^{14}$N+p	&	1	&	3/2	&	\textbf{-13.821$\times$10$^3$}\\
	&	&	&	$^{15}$O+$\gamma_{0.00}$	&	M1	&	1/2	&	\textbf{1.24}\\
     &	&	&	$^{15}$O+$\gamma_{6.79}$	&	E1	&	3/2	&	{-0.044}\\
%	&	&	&	$^{15}$O+$\gamma_{5.24}$	&	E1	&	5/2	&	{0.095}\\
%10.2817 & \textbf{10.2817}	&	5/2$^+$	&	$^{14}$N+p	&	2	&	3/2	&	\textbf{17.292$\times$10$^3$}\\
%	&	&	&	$^{15}$O+$\gamma_{6.79}$	&	M1	&	3/2	&	\textbf{0.2}\\	
%	&	&	&	$^{15}$O+$\gamma_{6.86}$	&	M1	&	5/2	&	\textbf{-0.4}\\	
%10.480 & \textbf{10.4675}	&	3/2$^-$	&	$^{14}$N+p	&	1	&	1/2	&	\textbf{28.998$\times$10$^3$}\\
%	&	&	&	$^{14}$N+p	&	1	&	3/2	&	\textbf{9.652$\times$10$^3$}\\
%	&	&	&	$^{15}$O+$\gamma_{0.00}$	&	M1	&	1/2	&	\textbf{-0.404}\\	
%	&	&	&	$^{15}$O+$\gamma_{6.79}$	&	E1	&	3/2	&	\textbf{0.1}\\
%	&	&	&	$^{15}$O+$\gamma_{6.86}$	&	E1	&	5/2	&	\textbf{0.1}\\
%10.506 & \textbf{10.5313}&	3/2$^+$	&	$^{14}$N+p	&	0	&	3/2	&	\textbf{205$\times$10$^3$}\\
%	&	&	&	$^{15}$O+$\gamma_{0.00}$	&	E1	&	1/2	&	\textbf{-0.195}\\
%	&	&	&	$^{15}$O+$\gamma_{6.79}$	&	M1	&	3/2	&	\textbf{0.3}\\
%	&	&	&	$^{15}$O+$\gamma_{6.86}$	&	M1	&	5/2	&	\textbf{-0.4}\\
%10.9288 & \textbf{10.9288}&	7/2$^+$	&	$^{14}$N+p	&	2	&	3/2	&	\textbf{56.948$\times$10$^3$}\\
%	&	&	&	$^{15}$O+$\gamma_{6.79}$	&	E2&	3/2	&	\textbf{1}\\
%11.218(3) & 11.217(2)& 3/2$^+$&	$^{14}$N+p	&	0	&	3/2	&	\textbf{40$\times$10$^3$}\\
%	&	&	&	$^{15}$O+$\gamma_{0.00}$	&	E1	&	1/2	& 5.21	\\   	
& {15}	&	3/2$^+$	&	$^{14}$N+p	&	0	&	3/2	&	\textbf{4.722$\times$10$^6$}\\
	&	&	&	$^{15}$O+$\gamma_{0.00}$	&	E1	&	1/2	&	\textbf{327.3}	\\
%& \textbf{15}	&	5/2$^+$	&	$^{14}$N+p	&	2	&	1/2	&	\textbf{1.452$\times$10$^7$}\\
\end{tabular}
\end{ruledtabular}
\label{table: fitParams}
\end{center}
\end{table*}

In examining the capture to the ground state in $^{15}$O, the $R$-matrix fits show the effect of our lifetime measurement. Specifically, in Fig.~\ref{fig: rmatrixRange}, we present fits showing the whole range of lifetimes for the 6.79 MeV state of $\tau = 0.6 \pm 0.4$. This shows that despite this measurement providing the most stringent limit on the lifetime, this range still translates to dramatic changes in the low energy behavior of the $S$-factor. Our fits, however, agree well with the capture data. One of the best fits using our lifetimes is shown alongside fits from Refs.~\cite{Runkle2005, Azuma2010, Adelberger2011, Li2016} in Fig.~\ref{fig: rmatrixClose}; the impact of higher energy data on the fit in the low energy range as presented in \citet{Li2016} will be addressed in a forthcoming paper that presents a broad range of new low energy cross section measurements. %By comparing the fits in these two Figures, the problems of finding a comprehensive fit at both low and high energies highlighted in \cite{Adelberger2011}, \cite{Li2016}, and \cite{Wagner2018} become apparent. In order to have a fit that achieves good agreement to the capture data at low energy, a systematic sacrifice must be made in the region of 1.0 - 1.5 MeV and vice versa.

\section{Summary and conclusions}
\label{sec: summary}

The $^{14}$N$(p,\gamma)^{15}$O reaction was used to populate excited states at 5.18 MeV, 6.17 MeV, and 6.79 MeV in $^{15}$O. The nitrogen targets were made by implantation on backings of Mo, Ta, and W. The Doppler shift of the $\gamma$-rays emitted by the decaying recoils were measured using the different targets and at seven different angles. A Monte Carlo simulation was applied to reproduce the experimental shifts and extract the lifetimes from the measured attenuation factors. By using multiple implanted targets of different backings, we were able to take a weighted average of our measurements to reduce the overall systematic uncertainty. Additionally, the Monte Carlo approach allowed us to recreate the depth profile of implanted targets with a high degree of accuracy, making the subsequent analysis based on the target composition more robust. The simulation also propagates uncertainties throughout every step, allowing it to reflect the experimental conditions more accurately. This is an improvement over previous measurements and their treatment of their targets.

The results show no evidence of systematic variations with previous measurements arising from the choice of backing materials. This work shows a larger uncertainty for the lifetime of the 5.18 MeV state but agrees within the uncertainties of the previous measurements. For the other transitions at 6.17 MeV and 6.79 MeV, the present measurement agrees well with the values reported by \cite{Bertone2001, Schurmann2008, Galinski2014}. Our work represents the only measurement of a finite lifetime for the 6.79 MeV state and provides even more stringent constraints on the level lifetimes. The discrepancies in previous measurements were resolved in this measurement with three different backings. These results, alongside recent measurements of the low-energy $^{14}$N$(p,\gamma)^{15}$O reaction capture cross-section will be used in a forthcoming paper to carry out a full $R$-Matrix fit over a very broad energy range to extrapolate the $S$-factor to astrophysical energies. This complete fit of and extrapolation from the data will allow for a more confident determination of this reaction's parameters and, therefore, a more complete understanding of this crucial reaction in the CNO cycle.

\begin{acknowledgments}
Fig.~\ref{fig: levelScheme} has been created using the SciDraw scientific figure preparation system [M. A. Caprio, Comput. Phys. Commun. 171, 107 (2005), http://scidraw.nd.edu]. This research utilized resources from the Notre Dame Center for Research Computing and was funded by the National Science Foundation through Grant No. PHY-2011890 (NSL), the Joint Institute for Nuclear Astrophysics - Center for the Evolution of the Elements Grant No. PHY-1430152, and the U.S. Department of Energy's (DOE) National Nuclear Security Administration (NNSA, Grant \# DE-NA0003888).
\end{acknowledgments}

\bibliography{lifetimesPaper}% Produces the bibliography via BibTeX.

\end{document}